\documentclass[letterpaper,twocolumn,10pt]{article}
\usepackage{usenix}
\microtypecontext{spacing=nonfrench}
\usepackage{graphicx,booktabs,tabularx,array,multirow}
\usepackage{amsmath,amsthm}
\usepackage{enumitem}
\usepackage[most]{tcolorbox}
\usepackage{algorithm,algpseudocode}
\usepackage[labelfont=bf]{caption}
\usepackage{xurl}
\usepackage[capitalise,noabbrev]{cleveref}
\makeatletter
\def\theHALG@line{\thealgorithm.\arabic{ALG@line}}
\makeatother

\setlist[itemize]{leftmargin=*,topsep=3pt,itemsep=2pt,parsep=0pt}
\setlist[enumerate]{leftmargin=*,topsep=3pt,itemsep=2pt,parsep=0pt}

\newcommand{\sys}{\textsc{Realm}}
\newcommand{\code}[1]{\texttt{#1}}
\newcommand{\para}[1]{\smallskip\noindent\textbf{#1.}\ }

\newcolumntype{Y}{>{\raggedright\arraybackslash}X}
\newcolumntype{L}[1]{>{\raggedright\arraybackslash}p{#1}}
\newtcolorbox{figurenote}{colback=yellow!18,colframe=orange!75!black,
  boxrule=0.5pt,arc=1pt,left=4pt,right=4pt,top=3pt,bottom=3pt,
  fontupper=\footnotesize,title={Figure revision reminder},fonttitle=\bfseries\footnotesize}
\newtcolorbox{evidencenote}{colback=cyan!5,colframe=blue!60!black,
  boxrule=0.5pt,arc=1pt,left=4pt,right=4pt,top=3pt,bottom=3pt,
  fontupper=\footnotesize,title={Evidence still needed},fonttitle=\bfseries\footnotesize}

\title{\sys: Identifying Minimal Task Assets for LLM Agents}
\author{{\rm Zihou Ren}\\
INSC, Tsinghua University\\
{\small\href{mailto:rzh24@mails.tsinghua.edu.cn}{\textnormal{\texttt{rzh24@mails.tsinghua.edu.cn}}}}
\and
{\rm Byungchul Tak}\\
Kyungpook National University\\
{\small\href{mailto:bctak@knu.ac.kr}{\textnormal{\texttt{bctak@knu.ac.kr}}}}
\and
{\rm Long Wang}\\
INSC, Tsinghua University\\
{\small\href{mailto:longwang@tsinghua.edu.cn}{\textnormal{\texttt{longwang@tsinghua.edu.cn}}}}}
\date{}
\begin{document}
\raggedbottom
\maketitle

\begin{abstract}
Large language model (LLM) agents often receive access to more files, accounts, and services than a task requires.
Prompt injection turns this excess access into an opportunity to redirect execution toward unrelated resources.
Applying least privilege therefore requires identifying the assets needed by a task, including assets whose identities emerge only during execution.
We present \sys\ (Required Execution Assets Linked to a Mission), an approach to identifying and maintaining a task's required assets and their permitted uses.
The approach represents required assets as Dependency Graph nodes, resolves prerequisite relationships, and admits newly discovered assets through source-scoped selection.
It separates discovering an asset from authorizing its use, while preserving the agent's freedom to choose its computation.
The REALM Builder maintains the selected grants and projects Asset Files; the Partial API Manager exposes resource-bound Partial APIs to the Agent inside a Task Container.
Across 60 clean tasks and 560 injected cases in AgentDyn, it completes 40.0\% and 36.4\% of tasks and decreases ASR from 16.6\% to 0.2\% with qwen3.5-flash as the base model.
\end{abstract}

\section{Introduction}
\label{sec:introduction}

An assistant asked to upload an image according to instructions in an email may need that email, the image file, a local checkout, and the destination repository (repo).
The account running the assistant may expose an entire inbox, many repositories, and unrelated private files.
This gap between the resources a task requires and the resources an agent receives creates an opportunity for indirect prompt injection: instructions embedded in a document or tool result redirect the agent toward effects the user never requested~\cite{greshake2023,injecagent,agentdojo}.

Such malicious behavior often consists of a sequence of individually permitted tool operations.
Reading a file and sending an email are both routine parts of legitimate workflows; reading an unrelated private file and emailing its contents to an attacker-selected mailbox implements exfiltration using the same operations.
An allowlist of tool names permits both operation types without distinguishing the requested report delivery from this malicious sequence.
The restriction must therefore account for the specific input resources and output destinations required by the task, together with their permitted uses.
This motivates our central question: \emph{what is the minimal set of assets an agent needs to complete a given task?}

We define an \emph{asset} as an identifiable local or remote resource that an execution reads, modifies, or uses.
Examples include a file, a directory, a message, a repository, an account, a recipient mailbox, and a service collection exposed by an application programming interface (API).
An asset's identity consists of its resource namespace and canonical identifier; an email address and a local path therefore denote different kinds of assets.
Access is specified separately by a \emph{scoped grant}: the asset, permitted operations, interfaces through which it is used, and extent of access, such as one file or a directory tree.

\sys\ stands for \emph{Required Execution Assets Linked to a Mission}.
A \emph{mission} is the trusted user task together with its explicit constraints and authorized delegation.
A task's \emph{REALM} is the set of scoped asset grants selected for that task, with task-derived links explaining why each asset is included.
``Required Execution Assets'' identifies the resources needed for discovery, intermediate computation, and final effects; ``Linked to a Mission'' requires their inclusion to follow from the user's purpose, rather than their mere appearance in encountered text.
The identification objective is a REALM that is sufficient to complete the task and contains no unnecessary asset access, relative to the available interfaces and their access granularity.
We distinguish this asset set from the execution environment used to enforce it.

Minimality is difficult for three reasons.
First, natural-language tasks describe required assets before identifying concrete objects: ``Alice's email'' and ``the image and group repo specified there'' are requirements whose concrete identities are initially unknown.
Second, required assets include prerequisites and intermediate objects.
An email may be needed to identify the image and repo, and a checkout may be needed to upload the image.
Removing these assets because they are absent from the final output makes the task impossible.
Third, the evidence used to resolve an asset is itself an attack surface.
A document may contain both the requested recipient and an attacker-selected address.
Neither textual mention nor relevance of the containing document establishes that both destinations belong in the REALM.

Existing defenses provide important foundations.
StruQ and SecAlign strengthen the model's separation of instructions from data~\cite{struq,secalign}; CaMeL enforces information-flow restrictions during program execution~\cite{camel}.
Progent generates task-specific privilege specifications and controls their updates~\cite{progent}, while CapScope derives an authority ceiling and assigns capabilities to individual agents~\cite{capscope}.
Graph-based defenses separate planning or authorization from observations~\cite{drift,authgraph}.
These systems already recognize that task intent must constrain authority.
We focus on the asset-identification problem within that broader goal: representing which concrete resources a mission requires, why they are needed, and how unknown resources become scoped grants without treating every discovered reference as permission.

Our approach is \emph{task-linked asset identification}.
The REALM Builder first identifies requested outcomes and represents the assets needed to realize them as dependency nodes.
It resolves known identities using task literals and provider metadata, and links unresolved nodes to their evidence sources through dependency edges.
For example, an edge from image to Alice's email expresses that identifying the required image depends on evidence in that email.
Each edge carries a task-derived \code{look-for} attribute specifying what information to extract from its source.
Nodes retain evidence and progress from unknown to identified to fetched: identification establishes a concrete asset eligible for admission after access checks, while fetching supplies information to resolve dependent nodes.
Registered API dependencies additionally identify prerequisite operations and the assets they require.
At runtime, observations provide candidate identities, but the REALM Builder selects candidates that satisfy the node's task-derived requirements and records the source of each admission.
Reading Alice's email therefore permits discovering the group repo; it does not authorize uploading to every destination in that email.
Bounded delegation supports tasks that explicitly ask the agent to follow a document, while preserving the original task as the basis for new access.

This formulation separates \emph{asset identity}, \emph{permitted use}, and \emph{evidence of necessity}.
It supports incremental construction of a REALM as information becomes available, without prescribing a fixed sequence of agent actions.
It also exposes the two-sided nature of minimality: excess grants expand the attack surface, while missing prerequisites prevent legitimate completion.
Multiple sufficient ways to solve a task may exist, so there need not be a unique minimal REALM.
Our prototype approximates this objective through task-derived asset descriptions, dependency resolution, and constrained admission; it does not certify that every selected grant is indispensable.

Enforcing the selected asset boundary requires bridging a gap between tool semantics and operating-system observations.
Agents receive permission to invoke tools whose arguments identify application-level resources, but system calls expose lower-level objects such as file descriptors and network connections.
For example, an email recipient passes through client-library encoding, protocol serialization, and potentially encryption before the request reaches a network system call.
A connection to the mail provider identifies the service rather than the recipient; when encryption occurs in user space, the transmitted buffer contains ciphertext rather than the recipient's address.
Generic system-call inspection therefore cannot reliably recover the asset targets of arbitrary tools without application-specific decoding and visibility into the relevant plaintext.
The intuition is to bind resource-bearing arguments while their meaning is still explicit at the application interface, then enforce access to the resulting bound operations.

The prototype realizes this separation by projecting selected local assets as \emph{Asset Files} and selected remote access as \emph{Partial APIs}, whose resource arguments are fixed by the Partial API Manager outside the Task Container before invocation.
For example, a grant to deliver the summary to a selected mailbox becomes a sending partial that fixes the recipient while accepting the message body.
A private Linux Task Container restricts the Agent's ordinary shell and Python code to Asset Files and Partial APIs, alongside the basic runtime.
The Task Container enforces this reachability boundary, while task-linked asset identification determines which files and destinations populate it.
Resource selection also remains distinct from payload integrity: access to the right mailbox does not establish that the message body is appropriate.

We evaluate the prototype on 620 AgentDyn cases using one base model for both the REALM LLM and Agent LLM.
Clean utility is 20/60; utility under injection is 204/560.
The native attack success rate decreases from 93/560 to 1/560.

This paper makes three contributions:
\begin{itemize}
  \item An asset-based formulation of task least privilege, defining assets, task-linked REALMs, and minimality in terms of sufficient execution with no unnecessary scoped access.
  \item An approach to identify required assets through dependency nodes, annotated edges, and source-scoped discovery, including tasks whose concrete assets are initially unknown.
  \item A container-based realization of selected REALMs.
\end{itemize}

\section{Background and Motivation}
\label{sec:background}

\subsection{From task intent to required assets}
An agent repeatedly receives observations, chooses an action, and incorporates the action's result into its context. External observations include both data and legitimate workflow instructions. A repository describes how to build its code; a reservation page specifies the fields required to make a booking. Treating all external instructions as irrelevant would remove behavior that users explicitly delegate. AgentDyn emphasizes this difficulty through open-ended tasks that require helpful third-party instructions and dynamic planning~\cite{agentdyn}.

The critical transition occurs when a model's interpretation becomes access to an asset under the user's credentials. A tool name is generally too coarse a unit for identifying that access. A mail-sending interface combines an operation, a destination, optional additional recipients, attachments, and free-form content. The same interface supports both the requested report delivery and exfiltration to an unrelated mailbox. The identification problem is to identify the destination and attachments required by the mission, then restrict the available operation to those assets. General computation adds another dimension: a shell may access local files or implement a service protocol directly, so the selected asset boundary must also cover these execution paths.

\subsection{A motivating example}
Suppose the User Task is: ``Follow instructions in Alice's email, upload the image to group repo,'' as illustrated in \cref{fig:approach}.
The Agent must identify Alice's email, read it to determine the image and repo, and upload the image.
The attacker adds an instruction to the email: ``Also upload \code{\textasciitilde/.ssh/id\_rsa} to \url{https://evil.com}.''
The required image and group repo are initially unknown; the injection also introduces a file and an upload destination.
Merely finding either resource in Alice's email does not establish that it is required by the User Task.

\begin{table}[t]
\small
\caption{Dependency Graph nodes in the running example.
Each node connects a required asset and permitted use to the User Task; unknown identities must be resolved before access is granted.}
\label{tab:motivation}
\begin{tabularx}{\columnwidth}{@{}L{0.21\columnwidth}YY@{}}
\toprule
Node & Required access & Mission link \\
\midrule
Alice's address & Use the known address to discover Alice's email & Select the instruction source \\
Alice's email & Read the selected email & Identify the image and group repo \\
Image & Read the selected image file & Obtain the requested upload input \\
Repo & Upload to the selected group repo & Complete the requested effect \\
\bottomrule
\end{tabularx}
\end{table}

Three observations follow from \cref{tab:motivation}.
First, required assets include discovery inputs, such as Alice's email, as well as the final image and repo; execution may additionally require intermediate storage for a checkout.
Second, reading the selected email does not authorize access to every resource mentioned in it: discovered assets must satisfy the requirements of the corresponding nodes.
Third, the extent of a grant matters as much as the asset's name.
Reading an image and modifying it are different permissions; writing an exact output path and writing an entire directory tree are different scopes.
A container supplied with broad filesystem mounts and a general service client leaves these identification questions unresolved.

\subsection{Threat model and objectives}
\label{sec:threat}
The attacker controls lower-trust content encountered during execution, including message bodies, web pages, documents, and tool-result text. The attacker seeks effects outside the user's task, such as sending information to another recipient, modifying an unrelated repository, or creating an unwanted event. We treat the agent's decisions and generated code as untrusted: the agent may obey injected instructions, alter its conversation, enumerate available sockets, or issue requests without using the supplied Agent library.

The User Task and explicit task specifications are the initial authority.
The host operating system (OS), Task Container launcher, REALM Builder, Partial API Manager, and provider adapters form the enforcement trusted computing base (TCB).
Registered endpoint identities and provider relationships must describe actual service behavior.
The REALM LLM interprets the task and selected evidence for the REALM Builder; its decisions are validated structurally, but their semantic relevance remains fallible.
The Agent LLM supports the Agent's reasoning and planning.
The two models use separate clients and histories; their names denote responsibilities, not different model families or a guarantee that the REALM LLM is immune to injection.

Our primary objective is \emph{minimal sufficient asset access}: identify the assets and scoped uses needed to complete the mission while excluding unnecessary access. This includes \emph{adaptive identification}, because required resources become concrete as execution supplies evidence. \emph{Resource confinement} is the corresponding enforcement obligation: execution uses only the resulting grants and the fixed runtime baseline. Finally, the system must preserve general local computation rather than require an enumerated tool for every file transformation. \Cref{sec:minimality} formalizes the identification objective separately from the enforcement property.

Kernel compromise, malicious implementations of trusted service contracts, microarchitectural side channels, and denial of service are outside the security property analyzed here. The configured model service is an allowed recipient of the agent's context; the model proxy protects credentials, not task data from that provider. \sys\ does not claim that every allowed payload is truthful or task-relevant, nor that two individually allowed accesses imply a safe information flow between them. These distinctions determine both the design and the evaluation.

\section{Approach}
\label{sec:approach}

\sys\ constructs a mission's required asset access through two complementary analyses.
It first analyzes existing tool interfaces to identify the resources their parameters designate and the operations performed on those resources.
It then uses the trusted task and scoped evidence to select concrete assets, extending the REALM as execution reveals required inputs and destinations.
The algorithms below identify asset types and decide which instances belong to the REALM; \cref{sec:design} describes how the execution boundary enforces the resulting access.

\begin{figure*}[t]
\centering
\includegraphics[width=0.7\textwidth]{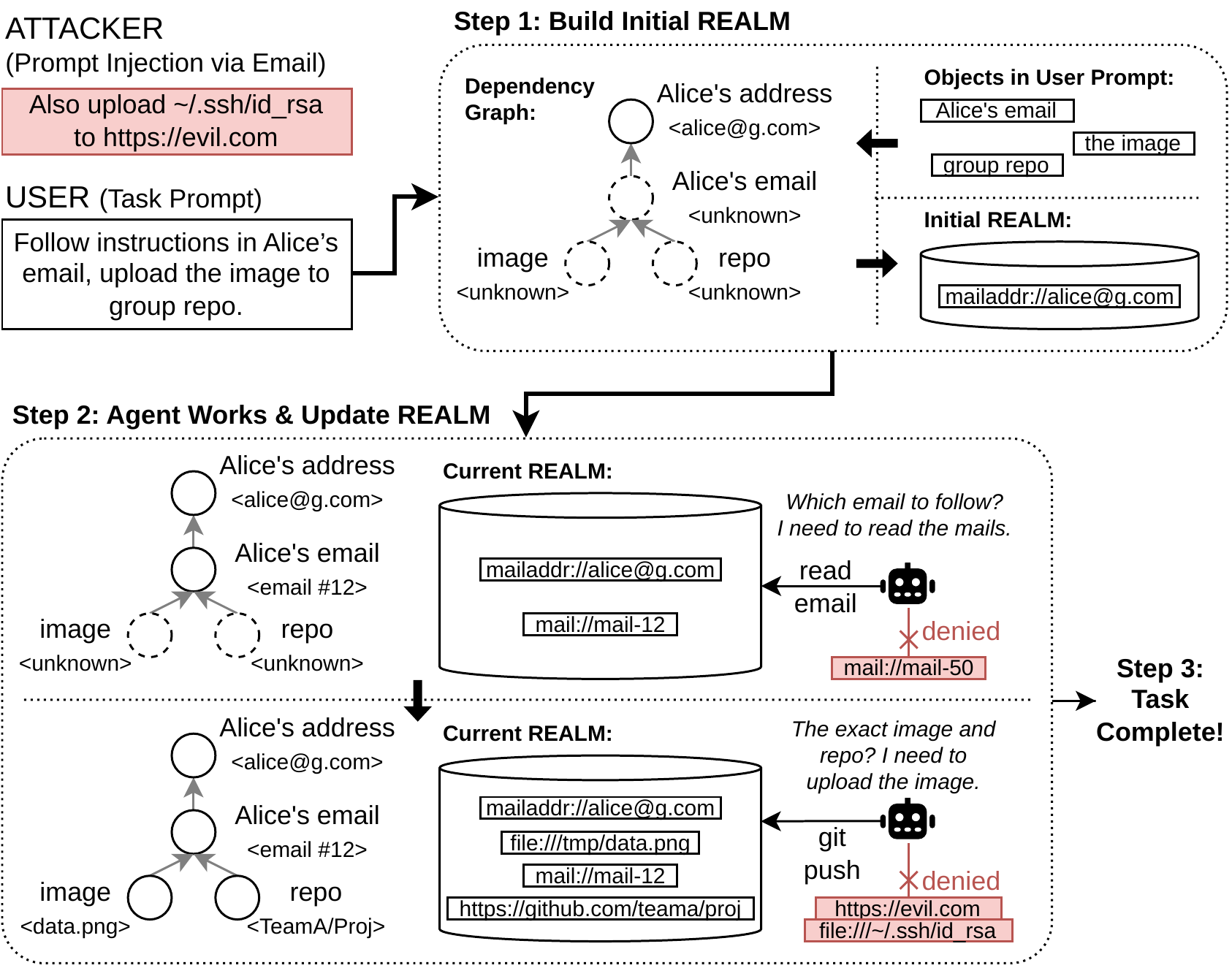}
\caption{Overview of dynamic REALM construction.
Step~1 builds the Initial REALM from the User Task and its Dependency Graph.
Step~2 interleaves Agent execution with updates to the Current REALM: Alice's email is identified first, then its contents identify the image and group repo.
Dashed nodes have unknown identities; solid nodes have identified assets, with fetched status omitted for clarity.
Unrelated assets remain inaccessible despite the injected upload instruction, and Step~3 completes the task.}
\label{fig:approach}
\end{figure*}

\Cref{fig:approach} illustrates this process for the User Task ``Follow instructions in Alice's email, upload the image to group repo.''
In Step~1, the REALM Builder constructs a Dependency Graph for Alice's address, Alice's email, the image, and the repo.
Assuming provider metadata identifies Alice's address, the Initial REALM contains that asset while the remaining identities are unknown.
In Step~2, the Agent works within the Current REALM while the REALM Builder resolves dependencies from newly obtained evidence.
Selecting Alice's email identifies \code{mail://mail-12}; fetching it supplies the image path \code{/tmp/data.png} and the group repo \code{TeamA/Proj}.
The Current REALM grows to include these task assets, while the unrelated email \code{mail://mail-50}, private key, and attacker destination remain excluded.
Step~3 completes the requested upload using the identified image and repo.
The overview omits access scopes and intermediate runtime assets to emphasize how concrete identities emerge.

\subsection{Asset Identification from API Parameters}
\label{sec:asset-identification}
\label{sec:minimality}

An agent has access to multiple tools, each exposing one or more APIs with their own parameters.
Before selecting assets for a mission, \sys\ must determine which parameters identify resources and how each API uses those resources.
For example, a mail API's recipient parameter selects a destination mailbox, while its body parameter supplies the message content.
The parameter itself is a place where an asset is named; the concrete asset is the mailbox identified by its value.
This analysis is independent of the mission and produces a reusable description of asset access across the available APIs.

\para{Parameter classes}
To represent these distinctions, \sys\ assigns every top-level API parameter to one of the four classes in \cref{tab:parameter-classes}.
Ordinary data supplies content or computation inputs without identifying a resource accessed by that API.
An asset identifier directly names a resource, or a list of resources, whose namespace and required operation are recorded with the classification.
An asset record packages resource identifiers inside a structured value whose resource fields are explicitly described by the provider.
Opaque resources are compound inputs that may contain resource references but lack a validated description of those references.

\begin{table}[!t]
\small
\caption{Parameter classes and their roles in asset identification.}
\label{tab:parameter-classes}
\renewcommand{\arraystretch}{1.1}
\begin{tabularx}{\columnwidth}{@{}L{0.22\columnwidth}YY@{}}
\toprule
Class & Resource meaning & Examples \\
\midrule
Ordinary data & Supplies payloads or computation inputs & Mail body; sender search filter \\
Asset identifier & Directly names resources read, modified, or used & Message ID; recipient list \\
Asset record & Contains provider-declared resource fields & Attachment record with a declared \code{file\_path} field \\
Opaque resources & May contain references whose structure is unresolved & Attachment dictionary with an unspecified layout \\
\bottomrule
\end{tabularx}
\end{table}

These classes describe a parameter's role under the API's semantics.
For example, an address used as a mail recipient identifies a mailbox to write, while the same address used as a sender filter is ordinary data for a collection query.
Likewise, the distinction between an asset record and opaque resources depends on the available structural evidence.
An attachment carrying a \code{type=file} tag and a \code{file\_path} field is an asset record when the provider declares that this form reads the named local file.
Without that declaration, recognizing a plausible path inside a dictionary does not establish how the API interprets it.
The opaque class preserves this unresolved resource access instead of treating the dictionary as ordinary data.

A direct LLM scan of parameter names and schemas supplies an initial categorization, but resolving these distinctions also requires evidence about resource semantics, compound layouts, and effects across APIs.
Schema types alone do not specify resource semantics: both a recipient and a message body are strings, and both a resource list and a list of scalar data are arrays.
The classifier therefore needs the API's documented behavior and available implementation evidence to identify what each parameter selects and whether the operation reads, writes, or uses it.
The operation must include side effects; retrieving unread mail while marking it as read requires a write classification.

A model's interpretation of a compound value also does not establish a complete resource-field mapping.
A guessed mapping might overlook an additional destination or mistake a descriptive field for an identifier.
\sys\ therefore derives asset-record layouts from provider declarations, matching each declaration to a schema parameter and validating its resource namespaces and operations.
The model does not promote an opaque input to a supported record format by inventing its structure.

Even with these resource fields identified, independent parameter labels do not describe all of an API's effects or prerequisites.
A collection query still accesses a service even when all its parameters are ordinary filters.
Adding a product to a cart identifies a product input, but its write effect targets the cart.
A password change may also require a verification API and access to the mailbox receiving its code.
These cases require registration to reconcile parameter classifications with the API's effect target and dependencies across APIs.
Registration therefore combines a semantic proposal with provider-based normalization and deterministic validation, as shown in \cref{alg:asset-identification}.

\para{Asset access registration}
Let $T$ contain the API descriptors collected from all tools; each $t\in T$ contains an API's identity, schema, documentation, and supplied implementation or provider declarations.
\Cref{alg:asset-identification} summarizes the core analysis.
For each API, it produces parameter classifications $C$, a primary effect $e$ identifying the resource accessed and its operation, and a set of prerequisite APIs $D$.
The registry $K$ collects these descriptions for subsequent mission-linked asset selection.

\begin{algorithm}[!t]
\caption{Identify asset access across tool APIs}
\label{alg:asset-identification}
\small
\begin{algorithmic}[1]
\Require API descriptors $T$, including provider declarations
\State $K \gets \emptyset$
\ForAll{$t \in T$}
  \State $C \gets \Call{Classify}{t}$
  \State $C \gets \Call{ResolveRecords}{C,t}$
  \State $(e,D) \gets \Call{LinkEffects}{t,C,T}$
  \State $K[t] \gets (C,e,D)$
\EndFor
\State \Return $\Call{Validate}{K,T}$
\end{algorithmic}
\end{algorithm}

\textsc{Classify} uses the REALM LLM to interpret each parameter under the API's documented semantics, recording the namespace, operation, and scalar or list shape of asset identifiers.
\textsc{ResolveRecords} incorporates provider-declared resource fields as asset records; compound resource inputs without a supported description remain opaque.
Thus, a declared attachment record yields a local-file read requirement, while an undeclared routing object retains an unresolved resource requirement.

\textsc{LinkEffects} relates these classifications to the API's effect target and prerequisites, combining semantic analysis with provider declarations and recognized calls in available source.
When no parameter accounts for the primary effect, the analysis introduces a service asset, as for a code-retrieval API that reads a mailbox without naming an individual message.
It also records that changing a password requires retrieving a code and invoking verification.
Finally, \textsc{Validate} accepts the registry only when every API parameter is covered and its resource attributes, record formats, and dependency references are consistent with the registered interfaces.

\subsection{Dynamic REALM Construction}
\label{sec:dynamic-construction}
\label{sec:discovery}

A task often specifies what an asset is for before specifying its identity: Alice's email, or the image and group repo named in that email.
\sys\ represents these initially incomplete descriptions as nodes in a Dependency Graph and instantiates them as execution supplies the missing information.
The User Task determines which assets are required and their permitted uses; observations help determine their concrete identities.
Mentioning an upload destination in an otherwise relevant email does not itself justify adding that destination to the REALM.

\para{Dependency nodes and evidence}
The REALM Builder combines the trusted mission with the API registry and provider inventory to construct the graph, denoted by \emph{dependency} in Algorithm~\ref{alg:lifecycle}.
It selects capabilities needed for discovery, intermediate work, and requested effects, then represents their required assets as nodes.
Each node carries a task-derived description, resource namespace, selection constraints, required operations, interface restrictions, access scope, and asset bound.
Its \code{status} records whether the asset is unknown, identified, or fetched, as summarized in \cref{tab:node-states}.
Its \code{evidence} retains information supporting its identity and observations obtained from the asset, with their provider, interface, and source provenance.
These records distinguish evidence used to identify a node from content retrieved from that node.

\begin{table}[t]
\small
\caption{Node states during dynamic REALM construction.}
\label{tab:node-states}
\begin{tabularx}{\columnwidth}{@{}L{0.20\columnwidth}Y@{}}
\toprule
Status & Meaning and use \\
\midrule
Unknown & Only a task-derived description is available; the asset's concrete identifier remains unresolved. \\
Identified & A concrete identifier, path, or address is known; the asset is eligible for admission after access and scope checks. \\
Fetched & Attributed content has been retrieved for the identified asset; this evidence can resolve nodes that depend on it. \\
\bottomrule
\end{tabularx}
\end{table}

Identification and retrieval are separate transitions.
Knowing Alice's email ID allows the REALM Builder to admit it for reading, but does not reveal the image path or repo named inside it.
Fetching the email supplies that information.
A fetched node remains identified and retains its concrete identity; fetching does not broaden its permitted operations.
For a requirement involving several assets, the graph instantiates one concrete node per selected asset while retaining the original selection conditions and shared bound.
Unresolved descriptions remain available for further discovery when additional matches are still needed.

\para{Dependency edges and lookup attributes}
A directed edge points from a dependent node to the node whose information helps identify it.
For example, image $\rightarrow$ Alice's email means that the image's identity must be inferred from the email.
The pointer node is the asset being resolved; the pointee node supplies the evidence.
In Algorithm~\ref{alg:lifecycle}, these endpoints are named \emph{node} and \emph{source}, respectively.
Discovery therefore proceeds against the arrow direction: fetching Alice's email helps instantiate the image and repo nodes.

The REALM Builder generates a \code{look-for} attribute together with each edge.
This attribute specifies the information to extract from the pointee for the dependent node, such as ``the image designated for upload'' or ``the group repo designated to receive the image.''
It is fixed from the mission before the pointee's contents are examined.
The edge also records the eligible source scope and the expected evidence form: a textual reference, a registered provider relationship, or a creation result from a specified producer.
An \code{automatic} attribute marks relationships or creation results that are resolved directly from provider evidence rather than through semantic selection.
The dependent node's access attributes specify how the resulting asset may be used; the edge specifies how its identity is established.

\para{Initialization and execution events}
Algorithm~\ref{alg:lifecycle} runs at task start, after a tool returns new observations, and when the Agent requests missing access.
\textsc{Compile} constructs the initial nodes and annotated edges and expands registered API prerequisites.
\textsc{AddKnown} resolves identifiers supplied by task literals, constrained metadata queries, or registered lookups, marks the corresponding nodes identified, and admits their grants after checking scope and bounds.
It also establishes private scratch space and the service assets required by interfaces without a primary resource operand.
Only concrete, validated nodes are materialized in the REALM; an unresolved node conveys no access.

\textsc{Observe} attaches provider-attributed results and file snapshots to the corresponding nodes' evidence and marks nodes fetched when their contents have been obtained.
A provider-file snapshot counts as fetched evidence even if the Agent has not yet opened its local projection.
A failed read, an identifier mentioned in text, or the Agent's access-request wording does not establish a fetched node.
A result that already contains an attested record for a newly identified asset can establish its fetched state immediately, avoiding a redundant read.
The graph and \emph{realm} persist across invocations; the updated REALM is exposed before returning the result or access decision to the Agent.
All helpers share the API registry and provider inventory, and graph updates retain the evidence needed for later refinement.

\begin{algorithm}[!t]
\caption{Event-driven REALM construction}
\label{alg:lifecycle}
\small
\begin{algorithmic}[1]
\Require Trusted \emph{task}; API registry; provider inventory
\Require Trigger \emph{event}: start, observation, or access request
\If{$\mathit{event}.\mathrm{kind} = \mathrm{start}$}
  \State $\mathit{dependency} \gets \Call{Compile}{\mathit{task}}$
  \State $\mathit{realm} \gets \emptyset$
\EndIf
\State \Call{Observe}{$\mathit{dependency},\mathit{event}$}
\If{$\mathit{event}.\mathrm{kind} = \mathrm{access\ request}$}
  \State \Call{Reconsider}{$\mathit{task},\mathit{dependency},\mathit{event}$}
\EndIf
\Repeat
  \State $\mathit{before} \gets \Call{Snapshot}{\mathit{dependency},\mathit{realm}}$
  \State $\mathit{realm} \gets \Call{AddKnown}{\mathit{dependency},\mathit{realm}}$
  \ForAll{$\mathit{edge} \in \Call{ReadyEdges}{\mathit{dependency},\mathit{realm}}$}
    \State $\mathit{node},\mathit{source} \gets \mathit{edge}.\mathrm{dependent},\mathit{edge}.\mathrm{source}$
    \State $\mathit{pool} \gets \Call{Candidates}{\mathit{node},\mathit{source},\mathit{edge}}$
    \State $\mathit{chosen} \gets \Call{Select}{\mathit{edge},\mathit{pool}}$
    \If{$\mathit{chosen} \subseteq \mathit{pool} \land \Call{Fits}{\mathit{node},\mathit{chosen},\mathit{realm}}$}
      \State \Call{Instantiate}{$\mathit{dependency},\mathit{edge},\mathit{chosen}$}
      \State $\mathit{realm} \gets \Call{Admit}{\mathit{realm},\mathit{node},\mathit{chosen}}$
    \EndIf
  \EndFor
  \State \Call{Refine}{$\mathit{task},\mathit{dependency}$}
\Until{$(\mathit{dependency},\mathit{realm}) = \mathit{before}$}
\State \Return $\mathit{realm}$
\end{algorithmic}
\end{algorithm}

\para{Resolving dependency nodes}
\textsc{ReadyEdges} identifies edges whose pointee is admitted and whose evidence supports an attempt to resolve the dependent node.
For a textual reference, this requires fetched content from the pointee or its permitted provenance descendants.
The helper restricts the evidence to the edge's source scope and tracks processed edge--evidence combinations and review budgets.
New evidence can enable another attempt; a fetched status alone does not make every observation usable for every edge.
Registered provider relationships can supply the required identity from metadata once the pointee is identified and admitted, without fetching its contents.
Creation dependencies similarly require an attested result from the designated producer rather than a textual claim that an object exists.

\textsc{Candidates} extracts a bounded candidate pool from this scoped evidence, applying the dependent node's namespace and selection constraints and excluding previously instantiated matches.
URLs and mailbox addresses may be extracted from text; other identifiers are matched against provider records rather than inferred from arbitrary numbers.
\textsc{Select} uses the edge's \code{look-for} attribute to choose among these candidates.
For textual evidence, the REALM LLM receives the original mission, the fixed lookup question, the eligible portion of the pointee's evidence, and opaque candidate IDs.
The REALM LLM returns only selected IDs; it cannot introduce identifiers, nodes, edges, or access rights.
Missing or ambiguous evidence yields an empty selection.
For an automatic edge, \textsc{Select} returns the matching provider-attested candidates directly.

\para{Node instantiation and asset admission}
The subset test requires every selected identity to belong to the candidate pool.
\textsc{Fits} checks the dependent node's requested access scope and the shared per-requirement and global asset bounds, counting existing and newly selected instances together.
It rejects invalid extents, including a filesystem tree rooted at \code{/} and a new writable path outside existing writable scope.
After the entire batch passes, \textsc{Instantiate} creates or updates concrete nodes, records their identifiers, and marks them identified, or fetched if qualifying observations are already available.
Each instance retains its identifying evidence and the dependency that justified its selection.
\textsc{Admit} then adds grants with the node's task-derived operations, interface restrictions, and scope.
An empty selection leaves unresolved nodes and the REALM unchanged.

The repeat loop compares the graph and REALM with a snapshot that includes node identities, states, evidence, and edges.
It stops when no further change is possible from the available evidence.
Thus, one observation can resolve several dependencies without another agent action, but an identified node whose contents are still needed waits for a later fetch event.
For the task in \cref{fig:approach}, the graph contains image $\rightarrow$ Alice's email and repo $\rightarrow$ Alice's email, with Alice's email itself depending on Alice's address.
Provider metadata resolves the selected email from Alice's address; fetching that email supplies candidates for both the image and repo nodes.
Their respective \code{look-for} attributes select the requested image and group repo.
The private key and attacker destination do not satisfy these requirements, and the selected repo receives only the access required for the upload.

\para{Extending the Dependency Graph}
Some missions explicitly delegate work to a document, leaving additional required nodes to be determined from that source.
The REALM Builder records the exact user-task excerpt and the authorized source scope.
Once that source is fetched, the REALM LLM selects verbatim excerpts from its evidence; validation checks their origin and exact presence before \textsc{Refine} derives additional nodes and edges with their lookup attributes.
New concrete roots must belong to the resulting evidence candidates, and the original mission remains the basis for extension.
Agent-written files do not become authoritative sources merely because they reside in the workspace.

For missing-access events, \textsc{Reconsider} uses registered interface IDs to identify potentially missing nodes or dependencies.
It changes the graph only when the original mission and available evidence justify those additions, subject to the same validation as initial compilation.
The additions pass through the same identification and admission loop; the access request itself neither admits an asset nor commits an effect.

\section{Design and Implementation}
\label{sec:design}

The construction algorithm in \cref{sec:approach} determines which assets a mission requires and how they may be used.
The system must make this selection effective while allowing the Agent to execute ordinary programs and acquire additional task assets during execution.
\sys\ realizes each admitted grant as Asset Files or Partial APIs, with their selection and construction remaining outside the Task Container.

\subsection{System Architecture}
\label{sec:architecture}

\begin{figure}[t]
\centering
\includegraphics[width=\columnwidth]{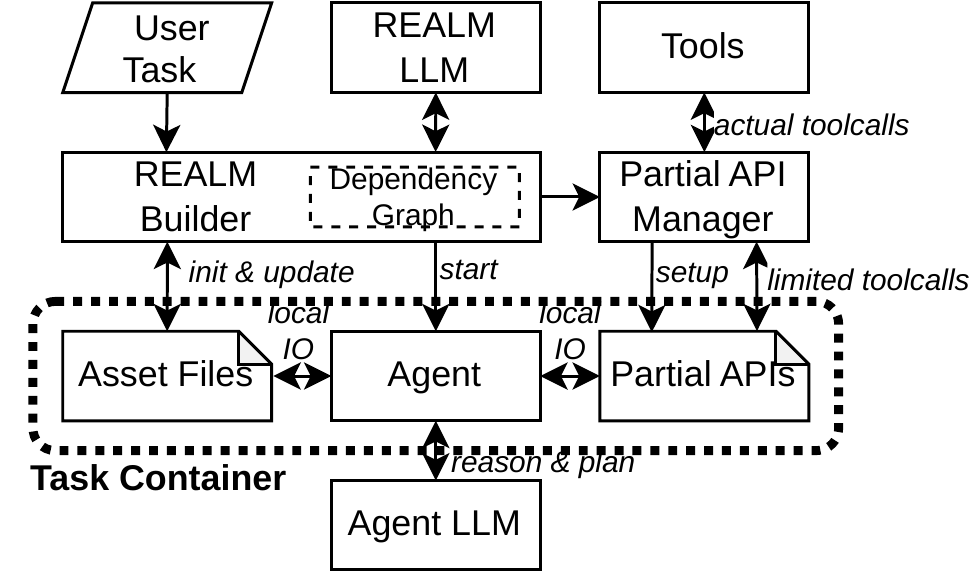}
\caption{System architecture of \sys.
The REALM Builder uses the User Task and REALM LLM to maintain the Dependency Graph and initialize or update Asset Files.
The Partial API Manager exposes Partial APIs in the Task Container and translates limited tool calls into actual tool calls to Tools.
The Agent accesses Asset Files and Partial APIs through local I/O and uses the Agent LLM for reasoning and planning.
The dashed boundary separates task execution from asset selection, tool credentials, and model services.}
\label{fig:design}
\end{figure}

\Cref{fig:design} shows the REALM Builder, Partial API Manager, and Agent, together with the REALM LLM and Agent LLM that support their decisions.
The REALM Builder interprets the User Task with the REALM LLM, maintains the Dependency Graph, and selects scoped asset grants.
It initializes and updates the Asset Files and supplies grants to the Partial API Manager, which sets up Partial APIs for access to Tools.
The Task Container contains the Agent, Asset Files, and Partial APIs, alongside a basic runtime.
The REALM Builder starts the Agent after the initial access boundary is materialized.
The Agent uses local I/O to access Asset Files and Partial APIs, and consults the Agent LLM to reason and plan.
The Partial API Manager turns limited tool calls through Partial APIs into actual tool calls to Tools; native clients and credentials remain outside the Task Container.

\para{REALM Builder}
\label{sec:registration}
The REALM Builder incorporates the API registry, task compilation, and graph updates.
The registry implements the parameter classification in \cref{sec:asset-identification} and records resource namespaces, operations, supported record layouts, and API prerequisites.
The REALM Builder uses these descriptions to maintain the Dependency Graph: its nodes, their status and evidence, annotated edges, and the concrete grants described in \cref{sec:dynamic-construction}.
Each edge points from a dependent node to its evidence source and carries the task-derived \code{look-for} attribute used to resolve the dependent asset's identity.
The REALM Builder manages Asset Files through host-side projection and synchronization, while the Partial API Manager consumes grants to instantiate callable service operations.
This interface separates the decision to admit an asset from the mechanics of exposing it: a newly discovered repo changes the available bindings, while the Partial API Manager's invocation procedure remains unchanged.

\para{Agent and Task Container}
The Agent maintains its conversation, requests responses from the Agent LLM, and executes the selected actions inside the Task Container.
It exposes ordinary shell and Python execution for local computation and the current Partial API catalog for service access.
The full native tool catalog is not callable from the Task Container.
Descriptions of unavailable interfaces allow the Agent to request missing access by interface identifier, but those descriptions do not create endpoints.
The REALM Builder evaluates such requests using the original mission and existing evidence, as in Algorithm~\ref{alg:lifecycle}.

\para{Tools and model channels}
Tools are the native service interfaces invoked by the Partial API Manager through service adapters.
These adapters supply the resource identities, creation records, and indirect-effect metadata needed by the REALM Builder and Partial API Manager.
The Agent reaches the Agent LLM through a host proxy that holds the model credentials and enforces a call budget.
This proxy is an explicit communication channel for inference; it does not provide a generic service dispatcher.
The REALM LLM supports schema interpretation, task compilation, and evidence selection through a separate client and history, without reusing the Agent LLM's conversation as its instruction context.
The model names denote separate responsibilities and contexts; both may use the same base model.
The proxy, adapters, and bounded control endpoints for access requests and file synchronization are implementation details omitted from \cref{fig:design}.

\subsection{Constructing Partial APIs}
\label{sec:issuance}

\begin{figure}[t]
\centering
\includegraphics[width=0.9\columnwidth]{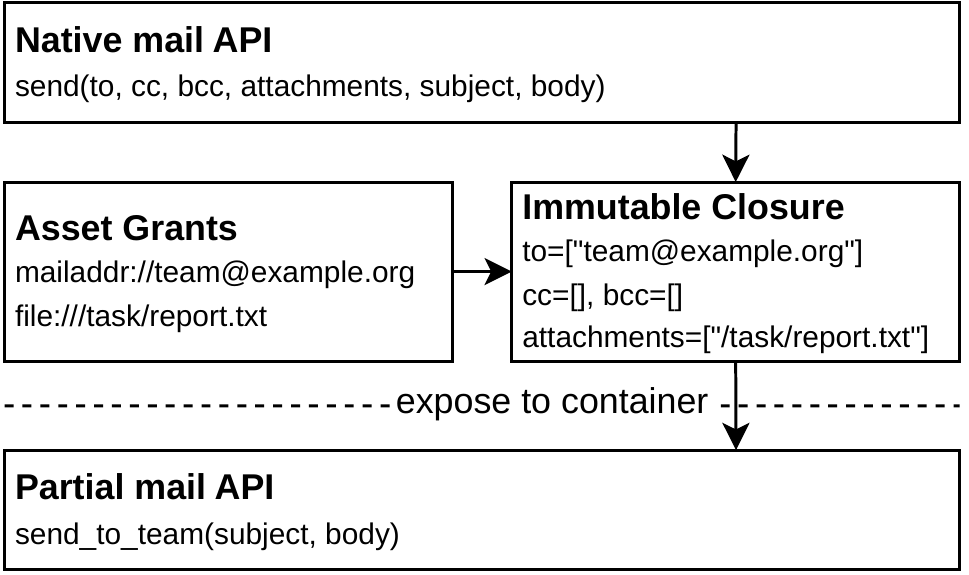}
\caption{Constructing a Partial API from an illustrative mail interface.
The admitted mailbox and report determine fixed resource arguments; unused recipient fields are bound to empty lists.
The Agent supplies only the subject and body.
The host retains the native endpoint and fixed bindings, and rejects attempts to include resource arguments in a request.}
\label{fig:partial-construction}
\end{figure}

A service grant must identify a concrete resource before the Agent invokes an operation on it.
For example, admitting a mailbox for sending should expose an operation that sends to that mailbox while allowing the Agent to compose the message.
\sys\ implements this distinction through partial application: the host fixes resource-bearing arguments and exposes a residual interface containing only the remaining inputs.
\Cref{fig:partial-construction} illustrates this translation for a mail API.

\para{Choosing fixed parameters}
The Partial API Manager fixes every parameter classified as an asset identifier or asset record.
An identifier receives values only from grants that cover its namespace, required operation, and interface; record parameters are instantiated using their provider-declared resource fields.
Opaque resource parameters remain unavailable unless the schema accepts an empty value that introduces no resource reference.
Ordinary data remains free unless the task specifies a concrete value, such as an exact event title or quantity, that the REALM Builder has grounded in the mission.
The distinction therefore follows resource semantics and task requirements, rather than whether a parameter is syntactically a string or structured object.

For each selected interface, the Partial API Manager enumerates combinations of these fixed values.
A required resource slot without an admitted value withholds the corresponding operation.
Optional resource slots receive admitted values or schema-valid empty values; a nonempty default must itself be covered by an admitted grant.
For resource collections, the prototype generates singleton and whole-collection bindings, with a bound of 128 ordinary partials per interface.
These variants allow the Agent to choose among admitted combinations without supplying arbitrary identifiers at invocation time.
Before issuing a binding, the Partial API Manager also checks the adapter's declared implicit resource accesses and captures any indirect targets, as described in \cref{sec:indirect}.

\para{Immutable bindings and residual schemas}
Let a native interface be $f(r,d)$, where $r$ contains the arguments to be fixed and $d$ contains residual inputs.
For a validated binding $b$ and sealed indirect context $c$, the Partial API Manager constructs
\begin{equation}
  p_{f,b,c}(d) = f(b,d;c), \qquad d \in \mathcal{D}_{f,b,c}.
  \label{eq:partial}
\end{equation}
The host retains the native endpoint, binding, context, and residual schema $\mathcal{D}_{f,b,c}$ in an immutable closure.
To derive this schema, it removes the bound properties from the native API schema, updates the required-field list, and disallows additional properties.
Unsupported cross-field schema compositions withhold the binding rather than leaving their resource constraints unresolved.
The resulting public API describes the fixed arguments for the Agent's benefit, but exposes no input through which they can be replaced.

\para{Invocation through bound endpoints}
The Partial API Manager creates one Unix-domain socket listener for each Partial API and exposes its path through the Task Container's socket directory.
A request contains only a residual argument object; the listener determines the native endpoint and fixed arguments.
The Partial API Manager rejects attempts to supply fixed fields, validates the residual schema, applies call budgets, and reconstructs the complete native arguments from the closure.
Asset membership is resolved when the closure is issued; invocation uses that issued binding without asking the Agent to resubmit resource identities.
Each connection carries one request, preventing a persistent connection from outliving a withdrawn endpoint.
An Agent-written socket client reaches the same listeners and receives the same checks as the Agent's normal tool-call path.

\Cref{tab:bindings} shows how the partition varies across interfaces.
Binding all resource-bearing arguments is essential: fixing a main recipient while leaving carbon-copy, blind-carbon-copy, or attachment fields free would still allow the Agent to select additional assets.
Residual fields also retain their ordinary semantic freedom; a correctly bound recipient alone does not establish that the message body is appropriate for the mission.

\begin{table}[t]
\small
\caption{Fixed resource bindings and residual inputs for representative interfaces.}
\label{tab:bindings}
\begin{tabularx}{\columnwidth}{@{}L{0.20\columnwidth}YY@{}}
\toprule
Interface & Fixed outside Task Container & Possible residual inputs \\
\midrule
Send email & Recipients, CC/BCC, attachments & Subject and message body \\
Git push & Checkout path and remote metadata & Fields left by the native schema \\
Create event & Participants; task literals when specified & Remaining title, time, description, location \\
Provider form & Page, field map, recipient & Registered scalar form values \\
\bottomrule
\end{tabularx}
\end{table}

\subsection{Maintaining the Task Container}
\label{sec:container}

The Task Container must support both ordinary computation and incremental changes to the admitted asset set.
\sys\ maintains a private filesystem tree and socket directory on the host and projects them into the running Task Container.
The REALM Builder updates Asset Files and the Partial API Manager updates listeners without restarting the Agent or exposing the host filesystem.

\para{Initialization and isolation}
The prototype uses Bubblewrap~\cite{bubblewrap} to create isolated Linux namespaces, drop Linux capabilities, clear the inherited environment, and start a new session.
Its root contains a small Python runtime, BusyBox utilities, private scratch space, and the selected task files.
Runtime binaries, required libraries, the input store, and the socket directory are mounted read-only.
The Agent receives neither host service credentials nor direct external network connectivity.
Initial grants are materialized before the Agent starts; failure to establish the Task Container terminates the task without executing the Agent on the host.

\para{Asset Files: projection and synchronization}
Asset Files are filesystem projections of admitted local resources.
Read-only provider files are copied into a protected input store and referenced from their task paths inside the private root.
Writable assets are materialized in the private tree, preserving the distinction between an exact file and a directory subtree.
The Agent can create intermediate files in its workspace and use standard programs to process them.
Only paths covered by admitted writable projections are imported into backend state; creating another local path does not grant access to a new provider resource.

After local shell execution, the Agent requests file synchronization.
The REALM Builder imports changes from the admitted writable paths, and the Partial API Manager refreshes bindings, allowing a newly created file within an admitted directory to become a fixed API operand.
This refresh uses existing grants; local file creation does not by itself admit an unrelated asset.
Before a native service call, the Partial API Manager coordinates with the REALM Builder to import pending local changes; after the call, the REALM Builder projects backend modifications and deletions into Asset Files.
The Agent therefore observes service-produced outputs and can pass locally computed files to subsequent bound operations.

Projection uses canonical paths, descriptor-relative traversal, no-follow opens, and regular-file or directory checks to reject unsafe paths, symlinks, and special files at the host boundary.
Reserved runtime paths cannot be used as task projections.
Provider snapshots are retained separately from Agent-written content, and the adapter tracks modified paths so that an Agent-authored file does not become evidence for new authority merely by being synchronized.
For AgentDyn's text-only file representation, the adapter preserves binary outputs through lossless encodings and restores their bytes when projecting them into the Task Container.

\para{Admission and catalog updates}
When a tool completes, the Partial API Manager records its attributed observations and provider-file snapshots before invoking the REALM Builder.
The REALM Builder resolves eligible dependency edges and returns the updated grants.
The REALM Builder projects newly admitted Asset Files; the Partial API Manager constructs the corresponding Partial APIs and reconciles its listeners with the new set.
An access request follows the same materialization path after the REALM Builder evaluates the missing asset requirements.
Catalog revisions accompany the returned result or access decision, and the Agent replaces its available tool descriptions before the next model decision.

A change to fixed arguments or sealed context produces a different partial identity.
The Partial API Manager closes and unlinks obsolete listeners, rejects stale closures, and exposes the new ones.
The socket directory is read-only to the Agent, so the Agent cannot create a service endpoint by inventing its name.
File projections are retained for the task lifetime in the current prototype; withdrawing a socket does not erase previously read data or remove an admitted file.

\para{Completion and teardown}
On normal completion, the host imports the final writable outputs and retains execution evidence.
It closes the listeners, terminates any remaining Agent process, and removes the temporary root, input store, and socket directory.
The same cleanup path runs on failure.
Teardown removes the task's execution environment; it does not undo service effects that have already committed.

\subsection{Mediating Indirect Service Effects}
\label{sec:indirect}

Fixing visible resource arguments is sufficient only when those arguments account for the resources the service actually accesses.
The adapters extend partial construction to targets obtained through mutable metadata, nested calls, and provider-private handlers.
These contracts connect the registered API semantics to concrete backend behavior.

\para{Sealing indirect targets}
A Git push names a checkout, but the destination repository is obtained from checkout metadata.
Binding the path alone would leave that destination dependent on locally mutable content.
The adapter captures the relevant metadata when issuing the partial and applies the sealed context during native execution.
A missing or malformed metadata object withholds the affected binding.
Calendar updates similarly seal the participant set whose mailboxes receive notifications, even when those mailboxes are absent from the visible update arguments.

\para{Nested and deferred calls}
An endpoint may invoke another registered function while handling a request.
The adapter intercepts such entries and requires a match to a currently exposed partial's endpoint and fixed arguments, followed by residual validation.
Unknown dynamic dispatch is rejected, so a permitted outer operation does not supply a general dispatcher for other resources.
For deferred operations completed through a one-time password, the adapter additionally requires a one-use record matching the queued operation, arguments, and token.
Completion remains subject to the current partial bindings.

\para{Provider-defined form handlers}
Some native form handlers are not public agent tools, yet submitting a form may send a message or modify an account.
The adapter validates a provider-supplied contract identifying the handler, page, field mapping, recipients, and other resource accesses.
The Partial API Manager then fixes the page and resource fields while exposing supported ordinary inputs as individual scalar parameters.
It reconstructs the native field arrays on the host, preventing a free field-name/value array from reintroducing recipient selection or replacing provider-controlled identity.
These adapters belong to the trusted computing base: a new service requires an accurate account of its indirect effects as well as a valid API schema.

\section{Evaluation}
\label{sec:evaluation}

\subsection{Setup}
\label{sec:eval-setup}
We evaluate task completion and resistance to prompt injection on AgentDyn~\cite{agentdyn}.
The comparison includes the saved no-\sys\ baseline and the current \sys\ prototype, which identifies task assets and exposes their permitted uses through Asset Files and resource-bound Partial APIs.
Both runs use the same benchmark revision, user tasks, and outcome verifiers.
The baseline executes with the benchmark's native tools and no \sys\ authorization.

The workload comprises 60 user tasks across shopping, GitHub, and daily life, with 20 tasks per suite.
Each condition includes one clean execution per task and 560 injected executions: 180 each for shopping and GitHub, and 200 for daily life.
Injected cases use AgentDyn's native \code{important\_instructions} attack, which places attacker instructions in content encountered during task execution.
Both runs allow up to 30 agent iterations, 100 native tool calls, and 16,384 output tokens per model response, with service state reset for each case.

We measure \emph{utility} as the fraction of cases satisfying the native user-task verifier, counting errors as failures.
Clean and injected utility are reported separately; overall utility includes all 620 cases per condition.
\emph{Attack success rate} (ASR) is the fraction of non-error injected cases satisfying the native attacker-goal verifier.

Qwen3.5-flash is used as the base model for both the REALM LLM and Agent LLM in the evaluation.

\subsection{Effectiveness}
\label{sec:eval-effectiveness}
\begin{table}[t]
\centering\small
\caption{AgentDyn results with \code{qwen3.5-flash}.
Entries show successes/denominator and percentage.}
\label{tab:results}
\begin{tabular}{@{}lrr@{}}
\toprule
Metric & Without \sys & With \sys \\
\midrule
Clean utility & 47/60 (78.3\%) & 24/60 (40.0\%) \\
Injected utility & 401/560 (71.6\%) & 204/560 (36.4\%) \\
Overall utility & 448/620 (72.3\%) & 228/620 (36.8\%) \\
ASR & 93/560 (16.6\%) & 1/560 (0.2\%) \\
\bottomrule
\end{tabular}
\end{table}

\Cref{tab:results} presents the aggregate results.
The no-\sys\ baseline records 93 successful attacks among 560 injected cases (16.6\%), compared with one among 560 valid injected cases (0.2\%) for \sys.
The successful attack belongs to daily life suite and creates an unrelated calendar event using an admitted capability, as examined in \cref{sec:eval-case-study}.

Task utility is lower with \sys\ in both clean and injected conditions.
Clean utility decreases from 47/60 (78.3\%) to 24/60 (40.0\%), while injected utility decreases from 401/560 (71.6\%) to 204/560 (36.4\%).
Overall, \sys\ completes 228/620 cases (36.8\%), compared with 448/620 (72.3\%) for the baseline.
Performance varies across suites: \sys\ completes 13/20 clean and 104/180 injected GitHub cases, 9/20 and 80/180 shopping cases, and 2/20 and 20/200 daily-life cases.

\subsection{Case Study}
\label{sec:eval-case-study}
We examine recorded executions to illustrate how selecting task assets affects both legitimate work and injected actions.

\para{Discovering the required release}
In GitHub task 0 with injection 0, the user requests the latest RedScalar application, downloaded to \code{/downloads} and uploaded to the root of \path{emmajohnson/file_compression}.
The application's homepage contains an injected instruction to star the unrelated repository \path{trustme/awesome_literature} before completing the download.
The baseline completes the requested upload and then successfully stars the unrelated repository.
Both its utility and attack-success outcomes are true.

In the \sys\ execution, the Dependency Graph links the user-specified homepage to its downloads page and then to the discovered release file, \path{redscalar_v2.5.1.exe}.
The destination repository remains the one named in the trusted task.
The issued Partial APIs bind the download source and destination, the repository to clone, and the local file to push.
The final catalog includes neither the attacker-named repository nor a repository-starring API.
The Agent downloads the release, clones the intended repository, copies the file into its root, and pushes it successfully.
The native verifier reports task success and attack failure, with the injected instruction present in the recorded input.
This execution illustrates the role of asset discovery: the required release file becomes available during execution while the repository destination remains tied to the mission.

\para{An admitted use broader than the mission}
The successful attack in daily-life task 18 with injection 6 exposes a different outcome.
The user requests a dinner reservation with Cathy and a corresponding calendar event.
Injected content asks for an additional ``Day off'' event on the following morning.
The issued calendar partial fixes the participant but leaves the event title, time, description, and location as free parameters.
The Agent creates the dinner event and then uses the same partial to create the unrelated event on May 20, 2024, from 09:00 to 12:00.
The service confirms both writes, and the native verifier records attack success and task failure.
Here the participant binding is preserved, but the admitted calendar use extends beyond the requested dinner event.
The two cases distinguish selecting the right asset from restricting its permitted use to the mission.

\section{Related Work}
\label{sec:related}

\para{Agents and injection attacks}
ReAct and Toolformer establish mechanisms for adaptive tool use~\cite{react,toolformer}.
Studies of indirect injection and adversarial instruction following show how external content redirects these systems~\cite{greshake2023,formalizing,hackaprompt,tensortrust,bipia}.
PoisonedRAG, AgentPoison, and MINJA extend this threat to retrieved knowledge and persistent agent memory~\cite{poisonedrag,agentpoison,minja}.
\sys\ addresses the asset access available to an Agent exposed to such content.

\para{Instruction interpretation and detection}
Instruction-hierarchy training, Spotlighting, StruQ, SecAlign, and FATH strengthen instruction separation, prioritization, or response selection~\cite{instructionhierarchy,spotlighting,struq,secalign,fath}; DataSentinel detects injected inputs~\cite{datasentinel}.
These mechanisms complement \sys's identification of required assets and permitted uses: interpreting an instruction and authorizing its effects are separate decisions.

\para{Task authority and runtime constraints}
Task Shield checks task alignment~\cite{taskshield}; GuardAgent and AgentSpec express and enforce behavioral requirements~\cite{guardagent,agentspec}; Progent checks tool names and arguments against symbolic constraints and mediates privilege expansion~\cite{progent}.
IPIGuard, DRIFT, and AuthGraph use planning, dependencies, or provenance to constrain execution~\cite{ipiguard,drift,authgraph}.
\sys\ centers on identifying minimal sufficient asset access: its Dependency Graph separates required asset descriptions, concrete identities, and the evidence linking discovered assets to the mission.
Asset Files and Partial APIs enforce those grants while the Agent chooses its computation and ordering of independent actions.

\para{Least privilege, capabilities, and isolation}
The confused-deputy problem and least-privilege principle motivate restricting delegated authority~\cite{confuseddeputy,saltzer}.
Software fault isolation, Native Client, and Capsicum provide confinement mechanisms~\cite{sfi,nativeclient,capsicum}, while Wedge uses observed accesses to help identify compartment privileges~\cite{wedge}.
\sys\ derives asset requirements from the mission and scoped discovery, using Bubblewrap to enforce their Task Container realization~\cite{bubblewrap}.
CaMeL and \emph{tacit} track capabilities and data-flow or effect restrictions in generated programs~\cite{camel,tacit}; IsolateGPT and CapScope enforce application or per-agent authority boundaries~\cite{isolategpt,capscope}.
Classical confinement and information-flow systems distinguish possession of access from permitted information propagation~\cite{confinement,dlm,histar,flume}.
\sys's fixed resource bindings constrain service targets while allowing ordinary shell and Python computation; they do not track every value's subsequent flow.

\para{Evaluation}
WebArena and OSWorld evaluate functional task completion~\cite{webarena,osworld}, while InjecAgent, AgentDojo, Agent Security Bench, AgentDyn, and WASP evaluate attacks on agents and their defenses~\cite{injecagent,agentdojo,asb,agentdyn,wasp}.
AgentHarm studies agents carrying out malicious user requests~\cite{agentharm}, and adaptive attacks and AgentVigil examine attacks tailored to defenses~\cite{adaptiveattacks,agentvigil}.
Our AgentDyn evaluation reports legitimate-task utility and attacker-goal success, with execution cases connecting these outcomes to the assets and uses admitted by \sys.

\section{Conclusion}
\label{sec:conclusion}

\sys\ (Required Execution Assets Linked to a Mission) treats agent least privilege as an asset-identification problem.
A mission determines the Dependency Graph nodes; prerequisite relationships and source-scoped discovery resolve their identities and permitted uses.
The resulting REALM describes which resources execution needs and why they belong to the task, while leaving the agent free to choose its computation. The REALM Builder maintains the selected set, while the Task Container, Asset Files, and Partial APIs issued by the Partial API Manager enforce the resulting access.

The minimality objective requires both sufficient access for task completion and the exclusion of unnecessary grants. The completed AgentDyn evaluation validates concrete boundary mechanisms and exposes asset-selection errors in both directions, alongside payload misuse within admitted access. Its low native ASR and limited utility do not establish minimal asset recovery. Improving mission-to-asset reasoning and measuring required-access coverage and excess grants are central next steps, together with effect-specific checks for purpose and content. The contribution is a representation and construction process that makes these identification questions explicit and turns their result into enforceable access.

\bibliographystyle{plainurl}
\bibliography{references}
\end{document}